\documentstyle[11pt,fullpage,doublespace,epsf]{article}
 \DeclareMathSizes{11}{19}{13}{9}   % For size 11 text

\makeatletter
\@addtoreset{equation}{section}
\makeatother

\begin{document}
  
\title{Cosmological Constant Seesaw\\ in
\\String/M-Theory}
\author{Michael McGuigan\\Brookhaven National Laboratory\\Upton NY 11973\\mcguigan@bnl.gov}
\date{}
\maketitle

\begin{abstract}In this paper we extend the Cosmological Constant Seesaw treatment of  hep-th/0602112 to String/M-Theory where the cosmological constant is finite. We discuss how transitions between different $\lambda$, one of Planckian vacuum energy, can give rise to a large $M_{Pl}^4$ denominator  in the Cosmological Constant Seesaw relation discussed by Banks, Motl and Carroll.  We apply these ideas to  2d/3d String/M-Theory and show how the existence of a large N dual fermionic theory makes the demonstration of a transition between different $\lambda$ relatively straight forward. We also consider  2d/3d Heterotic String/M-Theory cosmology, a theory for which the large N dual is unknown. The minisuperspace associated to these models  is 26/27 dimensional for the $SO(24)$ theory and 10/11 dimensional for the $SO(8) \times E_8$ theory and consists of the $T$ fields as well as the dilaton and metric. 2d Heterotic String Quantum Cosmology is similar to critical string dynamics except for the inclusion of the 2d gauge fields.  These 2d gauge fields have an important effect on the vacuum energy and on transitions between different $\lambda$ through the effects of Wilson lines. Finally we discuss the extension to existing higher dimensional string cosmologies possessing large N duals.
\end{abstract}
\clearpage

\section{Introduction}
In a previous paper \cite{McGuigan:2006hs} we studied the cosmological  constant seesaw relation \cite{Banks:2000fe}\cite{Motl}\cite{Carroll} $\lambda_- = \lambda_1^2/\lambda_2$ in the context of a quantized version of cosmology involving two universes with cosmological constants $\lambda_1$ and $\lambda_2$ by considering a generalized Wheeler-DeWitt (WDW) equation describing the two universes. Such descriptions are limited however by our current understanding of how to treat a mutiuniverse system in quantum gravity. Also in a low energy effective action the cosmological constant diverges so that one needs a UV complete theory which is valid at high energies to calculate the vacuum energy.

For the last point string/M-theory is an attractive possibility. In particular in string theory the vacuum energy can be represented as a  sum over world-sheets and is finite. In 2d string theory one can even compute the sum over genus and derive a closed expression valid to all orders in the coupling constant using a large N dual theory.  Finally in M-theory transitions between two string theories can be considered each of which can have different values of the cosmological constant. Because of these attractive features we will study the implementation of the cosmological constant seesaw in String/M-theory.

For point of reference consider the Friedmann equation :
\[
3M_{Pl}^2 \left( {\frac{{\dot a}}{a}} \right)^2  = \rho  - 3M_{Pl}^2 \frac{k}{{a^2 }} + \lambda _ -  
\]
where the reduced Planck mass is given by:
\[
M_{Pl}  = \frac{{m_{Pl} }}{{\sqrt {8\pi } }} = 4.340\mu g = 2.43534315 \times 10^{18} \frac{{GeV}}{{c^2 }}
\]
In this equation $a$ is the scale factor, $\rho$ is the matter-energy density, and $\lambda_-$ is the vacuum energy density or cosmological constant. The observed cosmological constant is of order:
\[
\lambda _ -   = \frac{{(2.4TeV)^8 }}{{M_{Pl}^4 }} =  10^{ - 120} M_{Pl}^4 
\]
in reduced Planck units. In terms of the ordinary Planck units 
\[
m_{Pl}  = \sqrt {\frac{{\hbar c}}{G}}  = 21.7645\mu g = 1.2209 \times 10^{19} \frac{{GeV}}{{c^2 }}
\]
the observed cosmological constant is given by (unreduced):
\[
\lambda _ -   = \frac{{(5.37TeV)^8 }}{{m_{Pl}^4 }} = 1.58 \times 10^{ - 122} m_{Pl}^4 
\]
The cosmological constant seesaw begins by forming the matrix:
\[
M^2  = \left( {\begin{array}{*{20}c}
   0 & {\lambda _1 }  \\
   {\lambda _1 } & {\lambda _2 }  \\
\end{array}} \right)
\]
with eigenvalues of magnitude:
\[
\begin{array}{l}
 \lambda _ -   \approx \frac{{\lambda _1^2 }}{{\lambda _2 }} \\ 
 \lambda _ +   \approx \lambda _2  \\ 
 \end{array}
\]
Then $\lambda_-$ is identified with the observed value of the cosmological constant today if $\lambda_1 = \sigma(2.4TeV)^4$ and $\lambda_2=\sigma^2 M_{Pl}^4$. What about the large eigenvalue $\lambda_+= \sigma^2 M_{Pl}^4 $? Solving the Friedmann equation for $\lambda = \lambda_+$ yields:
\[
a(t) = a(0)e^{t\sqrt {\lambda _ +  } /(M_{Pl} \sqrt 3 )}  = a(0)e^{t\sigma M_{Pl} /\sqrt 3 } 
\]
Thus assuming the early universe was described by a state with vacuum energy $\lambda_+$ one also has a period of inflation built into the cosmological constant seesaw in addition to estimate for the observed value now of $\lambda_- = \frac{{\lambda _1^2 }}{{\lambda _2 }}$. There are several remaining issues however. How can a universe transit from a state with vacuum energy $\lambda_+$ to one with vacuum energy $\lambda_-$? What is the interpretation of the matrix $M^2$? What is the origin of the large $M_{Pl}$ denominator factor? There are many possible approaches to these questions \cite{Banks:2000fe}\cite{Motl}\cite{Carroll}. One can consider perturbative expansion of the vacuum energy in $M_{Susy-Break}/M_{Pl}$ or the contribution to the vacuum energy of virtual black holes. The approach we took in \cite{McGuigan:2006hs} is to replace the Friedmann equation with the Wheeler-DeWitt (WDW) equation :
\[
 - \frac{1}{{12}}M_{Pl}^2 a^{ - 4} \frac{{\partial ^2 }}{{\partial a^2 }}\Psi (a) = (\rho  - 3M_{Pl}^2 \frac{k}{{a^2 }} + \lambda _ -  )\Psi (a)
\]
and use that as a starting point. In String/M-theory one has other potentially more robust methods using large $N$ dual Matrix or gauge theory descriptions of cosmology.  The point of this paper is to pursue the cosmological constant seesaw approach within string/M-theory where $\lambda_1$ and $\lambda_2$ are calculable, making use of a large $N$ dual description when available.

What about anthropic considerations? According to \cite{Denef:2006ad} the anthropic limit on the cosmological constant is \cite{Weinberg:1987dv}:
\[
 - 10^{ - 120} M_{Pl}^4  < \lambda _ -   < 10^{ - 118} M_{Pl}^4 
\]
The cosmological constant is in the anthropic window but we still need a mechanism to place it in the window. By analogy as the sky appears colored we can say that scattered light lies in the human visible spectrum but to understand why it's a particular color we need  to understand the scattering of light in the atmosphere. Likewise by observing spectral lines we can say energy transitions in atoms are in the human visible spectrum. To understand the form of the energy levels we need to solve Schrodinger's equation. Our approach to the cosmological constant seesaw is somewhat similar except we use the WDW equation  or a large $N$ matrix string/M cosmology instead of the Schrodinger equation. We know the cosmological constant is in the anthropic range but how does it attain it's value? If we can answer this question it should tell us a lot about the inner workings of quantum gravity.

This paper is organized as follows. In section 2 we set up the WDW equation for several 2d string theories and show how the calculated cosmological constant in these theories modifies the Friedmann equation. We also discuss the WDW equation associated with 3d M-theory, it's fermionic formulation, and how transitions between string models with different cosmological constants is relevant to the cosmological constant seesaw. In section 3 we repeat the same procedure for 2d/3d heterotic String/M-theory. Here we find a surprise: the $(a,\phi,T)$ space for these theories has the same dimension as critical bosonic and fermionic String and M-theory namely 26 and 27 for the bosonic case and 10 and 11 for  the fermionic case. The dimensions arise from the number of T fields in $SO(24)$ and $SO(8)\times E_8$ 2d heterotic strings together with fields associated with the metric and dilaton. Quantum cosmology in these models should be similar to  critical strings.  In section 4 we discuss the generalization to higher dimensions in particular to $R\times S^1\times S^2 \times K_6$, $R \times H^4 \times S^5$, and light cone Matrix cosmologies.

\section{Review of Cosmological Constant Seesaw}

We begin with a short review of the cosmological seesaw mechanism of \cite{McGuigan:2006hs} building on earlier discussion in \cite{Banks:2000fe}\cite{Motl}\cite{Carroll}

The cosmological constant seesaw relation $\lambda_- = \lambda_1^2/\lambda_2$
provides a phenomenologically interesting description of the observed cosmological constant. In this relation $\lambda_1$ is said to be generated by soft supersymmetry breaking \cite{Intriligator:2006dd}: 
\begin{equation}
\begin{array}{l}
 \lambda _1  = \frac{1}{{64\pi ^2 }}\sum\limits_i {( - )^{F_i } m_i^4 \log (m_i^2 /GeV^2 )}  \\ 
  \\ 
 \end{array}
\end{equation}
whereas $\lambda_2$ is generated by hard supersymmetry breaking \cite{Rohm:1983aq}:
\begin{equation}
\lambda _2  = \frac{1}{{64\pi ^2 }}M_{Pl}^4 \alpha ^2 
\end{equation}
Keith Dienes \cite{Dienes:2006ut} calculated $\alpha^2$ for a number of nonsupersymmetric heterotic tachyon free 4d string theories and found a range of values given by:
\[
\begin{array}{l}
 \alpha ^2  = (.0187, \ldots ,600) \\ 
 \alpha  = (.137, \ldots ,24.5) \\ 
 \end{array}
\]
Parameterizing the soft supersymmetry breaking vacuum energy as 
\[
 \lambda _1  = \frac{1}{{64\pi ^2 }}(5.37TeV)^4 \alpha  \\ 
 \]
we see that it is possible without fine tuning  to accommodate the observed vacuum energy 
\[
\lambda _ -   = (2.4TeV)^8 /M_{Pl}^4 
\]
using the cosmological seesaw relation $\lambda_-=\lambda_1^2/\lambda_2$.

Accommodating the observed vacuum energy is easy in the cosmological seesaw approach.
One simple extracts a pair of universes from the landscape, one with soft supersymmetry breaking and vacuum energy (2.1) and one with hard supersymmetry breaking and vacuum energy given by (2.2). The seesaw relation ensures that fine tuning is not required to get a realistic value of the cosmological constant. What is difficult is the interpretation. In \cite{McGuigan:2006hs} we interpreted the cosmological seesaw relation as coming from a modified Wheeler-DeWitt equation of the form:
\[
\begin{array}{l}
 \Delta \Phi _1  = \lambda _1 \Phi _1  + \sqrt {\lambda _1 \lambda _2 } \Phi _2  \\ 
 \Delta \Phi _2  = \sqrt {\lambda _1 \lambda _2 } \Phi _1  + (\lambda _1  + \lambda _2 )\Phi _2  \\ 
 \end{array}
\]
Here $\Delta$ is a second order (functional) differential operator. The equation can be solved by diagonalizing the right hand side with eigenvalues $\lambda_- \approx \lambda_1^2/\lambda_2$ and $\lambda_+ \approx \lambda_2$. Expanding in eigenfunctions $\Phi_-$ and $\Phi_+$
\[
\Phi_1 \approx \Phi_- - \frac{\lambda_1}{\lambda_2}\Phi_+
\]
and   $\Phi_1$ is identified with our universe evolving with cosmological constant  $\lambda_-$.

Another interpretation is even closer to the usual neutrino seesaw.
Recall the usual neutrino seesaw one has a Lagrangian of the form
\[
L = \bar \nu \partial \nu  + \bar N\partial N + m\bar \nu N + m\bar N\nu  + M\bar N_c N
\]
where the last term is a Majorana mass term for the right handed neutrino $N$ and $\nu$  is the usual left handed neutrino.
Taking our cue from this  and using the mapping:
\[
\begin{array}{l}
 \nu  \to \Phi _1  \\ 
 N \to \Phi _2  \\ 
 m \to \lambda _1  \\ 
 M \to \lambda _2  \\ 
 \partial  \to \Delta  \\ 
 \end{array}
\]
we write a Lagrangian as:
\[
L = \Phi _1^ +  \Delta \Phi _1  + \Phi _2^ +  \Delta \Phi _2  + \lambda _1 \Phi _1^ +  \Phi _2  + \lambda _1 \Phi _2^ +  \Phi _1  + \lambda _2 \Phi _2^ +  \Phi _2 
\]
This leads to WDW equations of the form
\[
\begin{array}{l}
 \Delta \Phi _1  = 0\Phi _1  + \lambda _1 \Phi _2  \\ 
 \Delta \Phi _2  = \lambda _1 \Phi _1  + \lambda _2 \Phi _2  \\ 
 \end{array}
\]
Diagonalizing using 
\[
U_L^ +  \left( {\begin{array}{*{20}c}
   0 & {\lambda _1 }  \\
   {\lambda _1 } & {\lambda _2 }  \\
\end{array}} \right)U_R  = U_L^ +  \left( {\begin{array}{*{20}c}
   0 & {\lambda _1 }  \\
   {\lambda _1 } & {\lambda _2 }  \\
\end{array}} \right)U_L \left( {\begin{array}{*{20}c}
   { - 1} & 0  \\
   0 & 1  \\
\end{array}} \right) \approx \left( {\begin{array}{*{20}c}
   {\lambda _1^2 /\lambda _2 } & 0  \\
   0 & {\lambda _2 }  \\
\end{array}} \right)
\]
and $\lambda _2  >  > \lambda _1$ we obtain the same eigenvalues as before. Here the interpretation is a little different as the initial vacuum energy associated with   $\Phi_1$ is zero and supersymmetry breaking is introduced through a mixing term $\lambda_1$ with the nonsupersymmetric theory  with large vacuum energy $\lambda_2$. The strength of the mixing term indicates the size of the supersymmetry breaking.

A simple way to think about the cosmological constant seesaw is that when two universes with different values of the cosmological constant interact we can model this by placing a source term on the right hand side of the Wheeler Dewitt equation for each of them. These equations follow from a Lagrangian which models the transition between the universes with different $\lambda$. "Integrating out"  the WDW field with $M_{Pl}^4$ vacuum energy yields a Lagrangian for a single universe whose effective cosmological constant has $M_{Pl}^4$ in the denominator as required by the cosmological constant seesaw relation. Indeed the relation $\lambda_{eff} =\frac {\lambda_1^2}{\lambda_2}$ follows from the simple integral relation:
\[
\int {D\Psi _2 } e^{i\lambda _1 (\Psi _1^+\Psi _2+ \Psi _2^+\Psi _1) +i\lambda _2 \Psi _2^ +  \Psi _2 } \sim  e^{ i \frac{\lambda _1^2 }{\lambda _2 }\Psi _1^ +  \Psi _1 } 
\]
In the above the novel physics is the transition between different $\lambda$ which cannot occur in the classical theory and is fundamental to the off diagonal terms in the coupled WDW equation. It is these transitions that we wish to study in the context of String/M-theory.

\section{2d/3d String/M-theory and transitions between different $\lambda$}

The cosmological constant seesaw relation $\lambda_-=\lambda_1^2/\lambda_2$ has the dimension of vacuum energy in any dimension (note that when $\lambda_1$ is not much smaller than $\lambda_2$ one uses a more complicated square root expression found in \cite{McGuigan:2006hs}). As the string landscape is simplest in 2d \cite{Seiberg:2005bx} it is reasonable to start there and then generalize to more complicated theories in higher dimensions. Also the vacuum energy can be computed in 2d string theory using matrix model techniques to all orders in the genus expansion and can serve as input into a WDW equation.

In this section we consider several 2d string theories and use the fermionic description and 3d M theory to demonstrate transitions between theories with different values of $\lambda$.

\subsection{0A 2d String Theory}

The low energy effective action for 2d 0A string theory is given by \cite{Seiberg:2005bx}:
\[
 S = \int {d^2 x(\sqrt { - g} } [e^{ - 2\phi } (R + 4(\nabla \phi )^2  + \frac{8}{{\alpha '}}) - \frac{1}{4}F^{(1)2}  - \frac{1}{4}F^{(2)2}  - \lambda _{0A} ] + L_T ) 
\]
\[ 
 L_T  = \sqrt { - g} e^{ - 2\phi } (\frac{1}{2}(\nabla T)^2  + \frac{1}{{2\alpha '}}T^2 ) + (\frac{1}{4}F^{(1)2}  - \frac{1}{4}F^{(2)2} )2T +  \ldots  \\ 
\]
Here $\phi$ is the dilaton, $F^{(1)}, F^{(2)}$ are two form fluxes and we have expanded to lowest order in the $T$ field. Parameterizing the metric as :
\[
ds^2  =  - (N^2  - a^{ - 2} N_1 N_1 )dt^2  + a^2 dx^2  + 2N_1 dxdt
\]
and varying the action with respect to $N$ yields the constraint \cite{Bilal:1993wm}:
\[
 - \frac{{e^{2\phi } }}{4a}\pi _a \pi _\phi   - \frac{{e^{2\phi } }}{4}\pi _a^2  + \frac{8}{{\alpha '}}e^{ - 2\phi }  + \frac{4}{{a^2 }}e^{ - 2\phi } ( - \phi '^2  - \frac{{a'}}{a}\phi ' + \phi '') = {\rho  + \frac{{\hat q^2 }}{4} + \lambda _{0A}  + \frac{1}{{a^2 }}\left( { - \frac{1}{{12}}} \right)} 
\]
where $\pi_a, \pi_{\phi}$ are canonical momentum. $\hat q = \left| q_1\right|+\left| q_2\right|$, and $q_1$, $q_2$ are the fluxes associated with $F^{(1)}$, $F^{(2)}$. There is type Bousso-Polchinski \cite{Bousso:2000xa} shift in the vacuum energy due to the fluxes. The last term is the Casimir energy associated to the $T$ field. $\rho$ is the energy density of the $T$ field.

Turning the canonical momentum into (functional) derivatives we obtain the WDW equation associated with the 0A Lagrangian:\[
\left( {\frac{{e^{2\phi } }}{4a}\partial _a \partial _\phi   + \frac{{e^{2\phi } }}{4}\partial _a^2  + \frac{8}{{\alpha '}}e^{ - 2\phi }  + \frac{4}{{a^2 }}e^{ - 2\phi } ( - \phi '^2  - \frac{{a'}}{a}\phi ' + \phi '')} \right)\Psi (a,\phi ,T)
\]
\[
 = \left( {\rho  + \frac{{\hat q^2 }}{4} + \lambda _{0A}  + \frac{1}{{a^2 }}\left( { - \frac{1}{{12}}} \right)} \right)\Psi (a,\phi ,T)
\]
The second order differential operator is associated with the Laplacian with respect to the following metric in $(a,\phi,T)$ space:
\[
\delta s^2  =  - 4a^2 e^{ - 2\phi } d\phi ^2  + 4ae^{ - 2\phi } dad\phi  + a^2 dT^2 
\]
The above equations are simplified if one assumes a homogeneous $R\times S^1$ cosmology. The 2d analog of the Friedmann equation then becomes:
\[
( - 4\frac{{\dot a}}{a}\dot \phi  + 4\dot \phi ^2  + \frac{8}{{\alpha '}})e^{ - 2\phi }  = \rho  + \frac{{\hat q^2 }}{4} + \lambda _{0A}  + \frac{1}{{a^2 }}\left( { - \frac{1}{{12}}} \right)
\]
The Hamiltonian constraint is then:
\[
 - \frac{{e^{2\phi } }}{{4a}}\pi _a \pi _\phi   - \frac{{e^{2\phi } }}{4}\pi _a^2  + e^{ - 2\phi } \frac{8}{{\alpha '}} = \rho  + \frac{{\hat q^2 }}{4} + \lambda _{0A}  + \frac{1}{{a^2 }}\left( { - \frac{1}{{12}}} \right)
\]
The WDW equation is written:
\[
 \left( {\frac{{e^{2\phi } }}{{4a}}\partial _a \partial _\phi   + \frac{{e^{2\phi } }}{4}\partial _a^2  + e^{ - 2\phi } \frac{8}{{\alpha '}}} \right)\Psi (a,\phi ) = \left( {\rho  + \frac{{\hat q^2 }}{4} + \lambda _{0A}  + \frac{1}{{a^2 }}\left( { - \frac{1}{{12}}} \right)} \right)\Psi (a,\phi ) \\ 
\]
If one assumes the $T$ field is weakly coupled the matter energy density can be expressed as:
 \[
\rho  = \frac{{N_L  + N_R }}{{a^2 }}
\]
where:
\[
\begin{array}{l}
 N_L  = \sum\limits_{n > 0} {N_L (n)\left| n \right|}  \\ 
 N_R  = \sum\limits_{n < 0} {N_R (n)\left| n \right|}  \\ 
 \end{array}
\]
are determined by left and right moving occupation numbers.

When the $T$ field is strongly coupled it is best to construct a large $N$ dual of the cosmology. This can be done along the lines of \cite{Karczmarek:2003pv} who introduced a time dependence of the large $N$ dual Hamiltonian condition determining the fermi surface of the theory.

\subsection{0B 2d String Theory}

The low energy effective action for 2d 0B string theory is given by \cite{Seiberg:2005bx}:
\[
 S = \int {d^2 x(\sqrt { - g} } [e^{ - 2\phi } (R + 4(\nabla \phi )^2  + \frac{8}{{\alpha '}}) - \frac{1}{4}\nu ^{(1)2}  - \frac{1}{4}\nu ^{(2)2}  - \lambda _{0B} ] + L_T  + L_C ) 
\]
\[
 L_C  = \sqrt { - g} ( - \frac{1}{2}(\nabla C)^2 ) + (\nabla C)^2 (T) +  \ldots )  
 \]
The the $\nu^{(1)}, \nu^{(2)}$ parameters are associated with the zero modes of the derivatives of $C$. Again parameterizing the metric as above we obtain the constraint:
\[
 - \frac{{e^{2\phi } }}{4a}\pi _a \pi _\phi   - \frac{{e^{2\phi } }}{4}\pi _a^2  + \frac{8}{{\alpha '}}e^{ - 2\phi }  + \frac{4}{{a^2 }}e^{ - 2\phi } ( - \phi '^2  - \frac{{a'}}{a}\phi ' + \phi '') = {\rho  + \frac{{\nu^2 }}{4} + \lambda _{0B}  + \frac{1}{{a^2 }}\left( { - \frac{2}{{12}}} \right)} 
\]
and the WDW equation associated with the 0B Lagrangian :
\[
\left( {\frac{{e^{2\phi } }}{4a}\partial _a \partial _\phi   + \frac{{e^{2\phi } }}{4}\partial _a^2  + \frac{8}{{\alpha '}}e^{ - 2\phi }  + \frac{4}{{a^2 }}e^{ - 2\phi } ( - \phi '^2  - \frac{{a'}}{a}\phi ' + \phi '')} \right)\Psi (a,\phi ,T, C)
\]
\[
 = \left( {\rho  + \frac{{\nu^2 }}{4} + \lambda _{0B}  + \frac{1}{{a^2 }}\left( { - \frac{2}{{12}}} \right)} \right)\Psi (a,\phi, T, C)
\]
The metric in $(a,\phi,T,C)$ space associated with the above Laplacian is:
\[
\delta s^2  =  - 4a^2 e^{ - 2\phi } d\phi ^2  + 4ae^{ - 2\phi } dad\phi  + a^2 e^{ - 2\phi } dT^2  + a^2 dC^2 
\]
The effective [potential for the $C$ field can be computed using the large $N$ dual theory \cite{Gross:2003zz}. It would be interesting to study the effect of this potential on the 2d 0B cosmology.

\subsection{IIA 2d String Theory}

The low energy effective action for 2d IIA string theory is given by \cite{Takayanagi:2004ge}:
\[
 S = \int {d^2 x(\sqrt { - g} } [e^{ - 2\phi } (R + 4(\nabla \phi )^2  + \frac{8}{{\alpha '}}) - \frac{1}{4}F^2  - \lambda _{IIA} ] + L_\Theta  ) 
\]
\[
 L_\Theta   = \sqrt { - g} (\bar \Theta D \Theta  + \bar \Theta F_{\mu \nu } \Gamma ^{\mu \nu } \Gamma ^3 \Theta ) \\ 
 \]
The constraint is given by
\[
 - \frac{{e^{2\phi } }}{4a}\pi _a \pi _\phi   - \frac{{e^{2\phi } }}{4}\pi _a^2  + \frac{8}{{\alpha '}}e^{ - 2\phi }  + \frac{4}{{a^2 }}e^{ - 2\phi } ( - \phi '^2  - \frac{{a'}}{a}\phi ' + \phi '') =  {\rho  + \frac{{q^2 }}{4} + \lambda _{IIA}  + \frac{1}{{a^2 }}\left( {\frac{2}{{12}}} \right)} 
\]
where $q$ is the single flux.
The WDW equation associated with the IIA Lagrangian is:
\[
\left( {\frac{{e^{2\phi } }}{4a}\partial _a \partial _\phi   + \frac{{e^{2\phi } }}{4}\partial _a^2  + \frac{8}{{\alpha '}}e^{ - 2\phi }  + \frac{4}{{a^2 }}e^{ - 2\phi } ( - \phi '^2  - \frac{{a'}}{a}\phi ' + \phi '')} \right)\Psi (a,\phi ,\Theta)
\]
\[
 = \left( {\rho  + \frac{{q^2 }}{4} + \lambda _{IIA}  + \frac{1}{{a^2 }}\left( {  \frac{2}{{12}}} \right)} \right)\Psi (a,\phi, \Theta)
\]
Again we have separated the Casimir term from the Hamiltonian of the matter field and included it as the last term above.

\subsection{IIB 2d String Theory}

The low energy effective action for 2d IIB string theory is given by \cite{Gukov:2003yp}:
\[
 S = \int {d^2 x(\sqrt { - g} } [e^{ - 2\phi } (R + 4(\nabla \phi )^2  + \frac{8}{{\alpha '}}) - \frac{1}{4}\nu ^2  - \lambda _{IIB} ] + L_{\Theta _ -  }  + L_{C_ +  } ) 
\]
\[
 L_{\Theta _ -  }  = \sqrt { - g} (\bar \Theta _{ - 1} D \Theta _{ - 1}  + \bar \Theta _{ - 2} D \Theta _{ - 2} ) \\ 
\]
The matter portion consists of two left moving fermions and one right moving boson. The parameter $\nu$ is related to the zero mode of the light cone derivative of $C$.
The constraint is given by:
\[
 - \frac{{e^{2\phi } }}{4a}\pi _a \pi _\phi   - \frac{{e^{2\phi } }}{4}\pi _a^2  + \frac{8}{{\alpha '}}e^{ - 2\phi }  + \frac{4}{{a^2 }}e^{ - 2\phi } ( - \phi '^2  - \frac{{a'}}{a}\phi ' + \phi '') =  {\rho  + \frac{{\nu^2 }}{4} + \lambda _{IIB}  + \frac{1}{{a^2 }}\left( {\frac{1}{{12}}} \right)} 
\]
The WDW equation associated with the IIB Lagrangian is:
\[
\left( {\frac{{e^{2\phi } }}{4a}\partial _a \partial _\phi   + \frac{{e^{2\phi } }}{4}\partial _a^2  + \frac{8}{{\alpha '}}e^{ - 2\phi }  + \frac{4}{{a^2 }}e^{ - 2\phi } ( - \phi '^2  - \frac{{a'}}{a}\phi ' + \phi '')} \right)\Psi (a,\phi ,\Theta _{ - 1} ,\Theta _{ - 2} ,C_+  ) 
\]
\[
 = \left( {\rho  + \frac{{\nu^2 }}{4} + \lambda _{IIB}  + \frac{1}{{a^2 }}\left( {  \frac{1}{{12}}} \right)} \right)\Psi (a,\phi ,\Theta _{ - 1} ,\Theta _{ - 2} ,C_ +  )
\]
The IIA and IIB 2d strings are related to the 0A and 0B strings by orbifolding and by interchanging holes with occupied states in the dual fermi sea picture \cite{Gukov:2003yp}.

\subsection{3d M-theory}

The low energy effective action for 3d M-theory is much less developed than for 2d string theory but is expected to be of the form \cite{McGuigan:2004sq}:
\[
S = \int {d^3 x(\sqrt { - g} } (\frac{1}{2}M_{Pl} R - \lambda _M  - \frac{1}{{24}}F_{\mu \nu \rho } F^{\mu \nu \rho }  + L_T )
\]
In the above $L_T$ is a Lagrangian for a single massless scalar field. The reason for including a single three  form flux $F$ is so the theory will have the same number of fluxes as the $0A$ 2d theory upon dimensional reduction. The three dimensional Newton constant is related to the reduced Planck Mass by $\frac{1}{{16\pi G}} = \frac{1}{2}M_{Pl}$.

Parameterizing the metric as:
\[
ds^2  =  - (N^2  + h^{ij} N_i N_j )dt^2  + h_{ij} dx^i dx^j 
\]
where
\[
h_{ij}  = V\frac{1}{{\tau _2 }}\left( {\begin{array}{*{20}c}
   {\tau _1^2  + \tau _2^2 } & {\tau _1 }  \\
   {\tau _1 } & 1  \\
\end{array}} \right)
\]
The two fluxes are given by:
\[
\begin{array}{l}
 \pi _1  = q_1  \\ 
 F_{\mu \nu \rho }  = q_2 \epsilon _{\mu \nu \rho }  \\ 
  \\ 
 \end{array}
\]
Varying with respect to $N$ we obtain the constraint:
\[
M_{Pl}^{ - 1} ( \pi _V^2  - \frac{1}{{V^2 }}m_2^2 (\pi _1^2  + \pi _2^2 )) + M_{Pl} R_h  - \rho  =  q_2^2  + \lambda _M  + \frac{{E(\tau )}}{{V^{3/2} }}
\]
In this expression we have separated out the Casimir energy of a single massless scalar field in 3d given by \cite{Seriu:1995hc}:
\[
E(\tau ) =  - \frac{{\tau _2^{3/2} }}{{4\pi }}\sum\limits_{n_1 ,n_2 \varepsilon Z/\{ 0,0\} } {\frac{1}{{|n_1  + \tau n_2 |^3 }}} 
\]

Turning the canonical momenta into operators we have the WDW equation associated with 3d M-theory.
\[
(M_{Pl}^{ - 1} ( - \partial _V^2  + \frac{1}{{V^2 }}\tau_2^2 (\partial_1^2  + \partial _2^2 )) + M_{Pl} R_h  - \rho)\Psi(V,\tau_1,\tau_2,T)
\]
\[
  =  (q_2^2  + \lambda _M  + \frac{{E(\tau )}}{{V^{3/2} }})\Psi(V,\tau_1,\tau_2,T)
\]
The second order differential operator is associated with the Laplacian with respect to the following metric in $V,\tau_1,\tau_2, T$ space:
\[
\delta s^2 = M_{Pl}(-dV^2 + V^2\tau_2^{-2}(d\tau_1^2 + d\tau_2^2)) + V^2dT^2
\]
Some simplification is possible if the cosmology is spatially flat and homogeneous. Then the analog of the Friedmann equation is:
\[
\frac{1}{4}M_{Pl} \frac{{\dot V^2 }}{{V^2 }} - \frac{1}{4}M_{Pl} \tau _2^{ - 2} (\dot \tau _1^2  + \dot \tau _2^2 ) - q_2^2  - \lambda_M  - \rho  = 0
\]
The Hamiltonian constraint is:
\[
M_{Pl}^{ - 1} \pi _V^2  - M_{Pl}^{ - 1} V^{ - 2} \tau _2^2 (\pi _1^2  + \pi _2^2 ) - q_2^2  - \lambda_M  - \rho  - EV^{ - 3/2}  = 0
\]
and the WDW equation becomes:
\[
\left( { - M_{Pl}^{ - 1} \partial _V^2  + M_{Pl}^{ - 1} V^{ - 2} \tau _2^2 (\partial _1^2  + \partial _2^2 )} \right)\Psi (a,\tau _1 ,\tau _2, T )
\]
\[
 = (q_2^2  + \lambda_M  + \rho  + EV^{ - 3/2} )\Psi (a,\tau _1 ,\tau _2, T )
\]

3d M-theory is supposed to be associated with 2d string theory by dimensional reduction (or generation). Following \cite{Horava:2005tt} we parametrize the 3d metric as
\[
ds^2  =  - N^2 dt^2  + a^2 e^{2\alpha \phi } dx^2  + \ell ^2 e^{ - 4\phi } (dy + A_1 dx)^2 
\]
The relation with previous variables is through:
\[
\begin{array}{l}
 V = a\ell e^{(\alpha  - 2)\phi }  \\ 
 \tau _1  = A_1  \\ 
 \tau _2  = a\ell ^{ - 1} e^{(\alpha  + 2)\phi }  \\ 
 \end{array}
\]
In terms of the new variables the analog of the Friedmann equation with zero flux is:
\[
( - 4\frac{{\dot a}}{a}\dot \phi  - 4\alpha \dot \phi ^2  - 2\lambda_M M_{Pl}^{ - 1} ) = 0
\]
For $\alpha = -1$ and $\frac{8}{\alpha^{\prime}} = - 2\lambda_M M_{Pl}^{ - 1} $ we obtain agreement with the 2d string low energy effective dilaton gravity. More complicated reductions from 3d gravity to 2d string dilaton gravity are considered in 
\cite{Achucarro:1992mb}\cite{Achucarro:1993fd}\cite{Cangemi:1992ri}.

\subsection{Fermionic formulation}
The WDW equations above can be used to describe the theories in an asymptotic regime of large volume and small coupling. One can use a fermionic formulation of the theory to describe a strongly interacting regime.

The fermionic formulation of the 0A theory is given by the 1+1 dimensional Hamiltonian:
\[
H_{0A}  = \int {dr} ( - \frac{1}{2}\partial _r \psi ^ +  \partial _r \psi  + ( - \frac{1}{2}\omega ^2 r^2  + \frac{{\hat q^2  - \frac{1}{4}}}{{r^2 }})\psi ^ +  \psi )
\]
 The fermi surface of the 0A theory is described by \cite{Maldacena:2005he}:
\[
\frac{1}{2}p_r^2  - \frac{1}{2}\omega ^2 r^2  + \frac{{\hat q^2  - \frac{1}{4}}}{{r^2 }} =  - \mu 
\]
where the chemical potential describes the coupling and $\omega^2 = 1/\alpha^{\prime}$
The fermionic formulation of the 0B theory is given by \cite{Maldacena:2005he}:
\[
H_{0B}  = \int {dx} ( - \frac{1}{2}\partial _x \psi ^ +  \partial _x \psi  + ( - \frac{1}{2}\omega ^2 x^2 )\psi ^ +  \psi )
\]
The fermi surface of the 0B theory is \cite{Maldacena:2005he}:
\[
\frac{1}{2}p_x^2  - \frac{1}{2}\omega ^2 x^2  =  - \mu 
\]
The relation of the fermionic field $\psi$ to the $T$ field is through the non local transform \cite{Martinec:2004td}:
\[
T(y,t) = \int{dxe^{yx} \psi^+(x,t) \psi(x,t)}
\]
In \cite{Horava:2005tt} a fermionic formulation of 3d M-theory was given by the Hamiltonian:
\[
H_M  = \int {dx} ( - \frac{1}{2}\partial _x \psi ^ +  \partial _x \psi  - \frac{1}{2}\partial _y \psi ^ +  \partial _y \psi  + ( - \frac{1}{2}\omega ^2 x^2  - \frac{1}{2}\omega ^2 y^2 )\psi ^ +  \psi )
\]
with fermi surface given by:
\[
\frac{1}{2}p_x^2  + \frac{1}{2}p_y^2  - \frac{1}{2}\omega ^2 x^2  - \frac{1}{2}\omega ^2 y^2  =  - \mu 
\]
Clearly the form of the above surface includes the 0A and 0B theory as special limits and thus $H_M$ forms an enveloping theory over the 2d string theories.

To include transitions between 0A and 0B theories one can introduce a time dependent fermi surface of the form:
\[
\frac{1}{2}p_r^2  - \frac{1}{2}\omega ^2 r^2  + \frac{{\hat q^2  - \frac{1}{4}}}{{r^2 }} - c(p_y  + y)e^{ - t}  =  - \mu 
\]
In terms of the 3d M-theory description we have:
\[
 \frac{1}{2}p_x^2  + \frac{1}{2}p_y^2  - \frac{1}{2}\omega ^2 x^2  - \frac{1}{2}\omega ^2 y^2  - c(p_y  + y)e^{ - t}  =  - \mu  
\]
\[
 yp_x  - xp_y  = \hat q \\ 
 \]
As the $(in)$ universe  and $(out)$ universe have different vacuum energy this modifies the right hand side of the WDW equation. This was the effect we were looking for: a transition between universes with different $ \lambda$. How about the other vacua in the 2d string landscape? According to \cite{Horava:2005tt} transition between IIA and IIB theories can also be constructed as orbifolded versions of the 0A and 0B theories. Thus transitions between those theories also seem possible within the fermionic formulation of 3d M-theory.

\subsection{Matrix theory formulation of 3d M-theory}

The fermionic formulation of 3d M-theory is not the only method to construct a transition between two different values of $\lambda$. A  Matrix theory associated with 3d M-theory was written in \cite{Park:2005pz} and is given by the following action:
\[
L = tr[\frac{1}{2}(D_t X)^2  + i\frac{1}{2}\psi D_t \psi  + X\psi \psi  + \frac{1}{2}\Lambda (t)X^2  + \rho_q (t)X]
\]
Here $X(t)$ ans $\psi$ are $N\times N $ real bosonic and fermion fields. $\Lambda(t)$ and $\rho_q(t)$ are deformation parameters associated with time dependent vacuum energy density and flux density in the theory. This theory is the dimensional reduction of Super Yang-Mills from two to one dimension with the inclusion of deformation. One can look at this Lagrangian as an alternative to the fermionic description of 3d M-theory. As a large $N$ Matrix description it is somewhat closer to higher dimensional approaches. If $\Lambda(t)$ take different values in the $(in)$ and $(out)$ state we can similarly use the Lagrangian to effect the transition between different values of $\lambda$ using the large $N$ dual theory. 

\section{2d/3d Heterotic String/M theory}

In this section we consider 2d/3d heterotic cosmologies which are much more complex than those of the previous section.

\subsection{SO(24) 2d Heterotic string theory}

The low energy effective action of SO(24) 2d Heterotic string theory is given by \cite{McGuigan:1991qp}\cite{Davis:2005qe}:
\[
S = \int {d^2 x\sqrt { - g} } e^{ - 2\phi } (R + 4\left( {\nabla \phi } \right)^2  + \frac{8}{{\alpha '}} - DT^A DT^A  - \frac{1}{4}F^2 )
\]
where $T^A$ are a set of 24 massless scalars.

Writing the metric as before and varying with respect to $N$ we obtain the constraint:
\[
 - \frac{{e^{2\phi } }}{4}\pi _a \pi _\phi   - \frac{{e^{2\phi } }}{4}\pi _a^2  + \frac{8}{{\alpha '}}e^{ - 2\phi }  + \frac{4}{{a^2 }}e^{ - 2\phi } ( - \phi '^2  - \frac{{a'}}{a}\phi ' + \phi '') =  {\rho  + \frac{{\nu^2 }}{4} + \lambda _{HO}  + \frac{1}{{a^2 }}\left( {-\frac{24}{{12}}} \right)} 
\]
Turning canonical momentum into functional derivative operators we have the WDW equation for 2d $SO(24)$ Heterotic String theory.
\[
\left( {\frac{{e^{2\phi } }}{4}\partial _a \partial _\phi   + \frac{{e^{2\phi } }}{4}\partial _a^2  + \frac{8}{{\alpha '}}e^{ - 2\phi }  + \frac{4}{{a^2 }}e^{ - 2\phi } ( - \phi '^2  - \frac{{a'}}{a}\phi ' + \phi '')} \right)\Psi (a,\phi ,T,A)
\]
\[
 = \left( {\rho  + \frac{{\nu^2 }}{4} + \lambda _{HO}  + \frac{1}{{a^2 }}\left( { - \frac{24}{{12}}} \right)} \right)\Psi (a,\phi, T,A)
\]
The second derivative operator is a Laplacian associated with the metric
\[
\delta s^2  =  - 4a^2 e^{ - 2\phi } d\phi ^2  + 4ae^{ - 2\phi } dad\phi  + \sum\limits_{A = 1}^{24} {a^2 e^{ - 2\phi } dT^{A2} } 
\]
The configuration space $(a,\phi,T^A)$ is the same dimension as the critical bosonic string with 24 variables coming from the $T $ fields and the remaining  2 coming from the $a$ and $\phi$ fields.

\subsection{$SO(8)\times E_8$ 2d Heterotic String theory}

The low energy effective action of $SO(8)\times E_8$ 2d Heterotic string theory is given by \cite{McGuigan:1991qp}\cite{Davis:2005qe}:
\[
S = \int {d^2 x\sqrt { - g(} } e^{ - 2\phi } (R + 4\left( {\nabla \phi } \right)^2  + \frac{8}{{\alpha '}} - DT^A DT^A  + \bar \Theta _ - ^\alpha  D\Theta _ - ^\alpha   + \bar \Theta _ + ^{\dot \alpha } D\Theta _ + ^{\dot \alpha }  - \frac{1}{4}F^2)
\]
where $T^A$ are 8 massless scalars and $\Theta_+$  are 8 right moving fermions in spinor representation of $SO(8)$ and $\Theta_+$ are 8 left moving fermions in conjugate spinor representation of $SO(8)$. 

Varying with respect to $N$ we obtain the constraint:
\[
 - \frac{{e^{2\phi } }}{4}\pi _a \pi _\phi   - \frac{{e^{2\phi } }}{4}\pi _a^2  + \frac{8}{{\alpha '}}e^{ - 2\phi }  + \frac{4}{{a^2 }}e^{ - 2\phi } ( - \phi '^2  - \frac{{a'}}{a}\phi ' + \phi '') =  {\rho  +  \lambda _{HE}  + \frac{1}{{a^2 }}\left( {\frac{0}{{12}}} \right)} 
\]
Turning canonical momentum into functional derivative operators we have the WDW equation for the 2d $SO(8)\times E_8$ Heterotic String theory.
\[
\left( {\frac{{e^{2\phi } }}{4}\partial _a \partial _\phi   + \frac{{e^{2\phi } }}{4}\partial _a^2  + \frac{8}{{\alpha '}}e^{ - 2\phi }  + \frac{4}{{a^2 }}e^{ - 2\phi } ( - \phi '^2  - \frac{{a'}}{a}\phi ' + \phi '')} \right)\Psi (a,\phi ,T,\Theta,A)
\]
\[
 = \left( {\rho  + \frac{{\nu^2 }}{4} + \lambda _{HE}  + \frac{1}{{a^2 }}\left( {  \frac{0}{{12}}} \right)} \right)\Psi (a,\phi, T, \Theta, A)
\]
The second derivative operator is a Laplacian associated with the metric
\[
\delta s^2  =  - 4a^2 e^{ - 2\phi } d\phi ^2  + 4ae^{ - 2\phi } dad\phi  + \sum\limits_{A = 1}^8 {a^2 e^{ - 2\phi } dT^{A2}} 
\]
The configuration space $(a,\phi,T^A)$ is the same dimension as the critical fermionic string with 8 variables coming from the $T^A$ fields and the remaining  2 coming from the $a$ and $\phi$ fields.

\subsection{3d Heterotic M-theory}

A plausible low energy effective action for 3d Heterotic M-theory is \cite{McGuigan:2004sq}:
\[
S = \int\limits_M {d^3 x(\sqrt { - g} (\frac{1}{2}M_{Pl} R - \lambda - \frac{1}{{24}}F_{\mu \nu \rho } F^{\mu \nu \rho } )}
\]
\[
  + \int\limits_{\partial _1 M} {d^2 x(L_T  + L_\Theta   + F_{SO(8)}^2 )}  + \int\limits_{\partial _2 M} {d^2 x(F_{E8}^2 )} 
\]
defined on the 3d cylinder $R\times [0,1] \times S^1$ and bounded by two hypersurfaces which contain the $SO(8)$ and $E_8$ fields separately. Type I/Heterotic string duality would then yield a 3d M-theory description of the $SO(24)$ Heterotic string.

One can proceed with the canonical formalism as in the previous section. This represents a 2+1 version of the Horava-Witten model. The dimension of the configuration space for the 3d theory is $(h_{ij},T^A)$ which is 27 dimensional for $SO(24)$ theory and 11 dimensional fro $SO(8)\times E_8$ theory. This is the same as the dimension of critical bosonic and fermionic M-theory respectively.

\subsection{Wilson Lines}

No matrix model or fermionic formulation has been constructed for the 2d Heterotic string. An argument for it's existence can be based on type I/ Heterotic duality in analogy with 10 dimensions \cite{Polchinski:1995df}. The fact that SO(32) strings can be open \cite{Polchinski:2005bg} suggests an even closer connection associated with long strings which are known to be present for the 2d SO(24) string \cite{Seiberg:2005nk}.

A 2d type I Matrix model was constructed by Minihan \cite{Minahan:1992bz} through introducing boundaries to discretized world sheets, soaking up free vertices with fermionic coordinates and then integrating these out. He found:
\[
H = \sum\limits_a { - \frac{1}{2}} \frac{{\partial ^2 }}{{\partial x_a^2 }} + ( - \frac{1}{2}\omega ^2 x_a^2 ) - 2(S_{za}  - \frac{1}{2})(\gamma x_a^2  - \mu ) + \sum\limits_{a < b} {\frac{{1/4 - \vec S_a  \cdot \vec S_b }}{{(x_a  - x_b )^2 }}} 
\]
The presence of the spin Hamiltonian is intriguing and because it is also present in the dual of type II string on $AdS_5 \times S^5$ \cite{Beisert:2003xu}. There the generalization of spin to that of a larger supergroup was essential. Extending the spin chain to the $SO(24)$ group could in principle introduce the $SO(24)$ gauge group into the matrix model.

10d Heterotic Matrix String theory has been constructed. This involves a 2d $O(N)$ gauge theory with 8  $X^i$ and $\theta_s$ fields transforming as a symmetric tensor under $O(N)$, 8 $\theta_c$ fields  transforming as an antisymmetric tensor under $O(N)$ and 32 left moving fermions $\chi$ transforming in the vector representation of $O(N)$. The action for the large $N$ dual is then \cite{Banks:1997it}\cite{Krogh:1998rw}\cite{Krogh:1998vb}\cite{Bonelli:1999pq}:
\[
S = \frac{1}{\pi }\int {d^2 w\{ Tr(} D_w X^i D_{\bar w} X^i  + \frac{1}{{4g^2 }}F_{w\bar w}^2  - \frac{{g^2 }}{2}[X^i ,X^j ]^2 
\]
\[
 + i(\theta _s D_{\bar w} \theta _s  - \theta _c D_w\theta _c  
 - 2g\theta _s \Gamma _i [X^i ,\theta _c ]) + i\chi D_w \chi \} 
\]
The challenge is to find a similar construction for the 2d heterotic theories.

In lieu of these developments one can still contemplate transitions between $HE$ and $HO$ by the introduction of Wilson lines \cite{Davis:2005qe}\cite{Davis:2005qi}. After compactification on a circle it is known that turning on Wilson lines allows one to continuously transform from the $HE$ to $HO$.  The vacuum energy of the $SO(24)$ 2d heterotic theory can be written:
\[
\rho  = \lambda _{HO}  + \frac{b}{{a^2 }}
\]
where
\[
b =  - (2 + \sum\limits_{i = 1}^{12} {(A_i^2  - |A_i |} ))
\]
The vacuum energy for the $SO(8)\times E_8$ 2d heterotic theory is given by: \cite{Davis:2005qi}:
\[
\rho  = \lambda _{HE}  + \frac{b}{{a^2 }}
\]
where in this case \cite{Davis:2005qi}:
\[
b =  - ( - \sum\limits_{i = 1}^4 {|A_i |}  + \frac{1}{8}\sum\limits_\varepsilon  | \sum\limits_{i = 1}^4 {\varepsilon _i A_i |} )
\]
For the special choice of Wilson lines \cite{Davis:2005qe}\cite{Davis:2005qi}:
\[
A_{HO}  = (0^4 ,(\frac{1}{2})^8 )
\]
\[
A_{HE}  = (1,0,1,0)
\]
the spectrum and vacuum energies of the $HO$ and $HE$ theories matches so that the theories can be transformed into one another continuously by varying the values of the $A_i$. As the $SO(24)$ and $SO(8)\times E_8$ theories have different vacuum energy this amounts to the  transition between different values of $\lambda$ that we were seeking.

Finally the 2d Heterotic string is conjectured to be dual to $N= (2,1)$ string \cite{Giveon:2004zz} and the 3d M-theory to the topological A model \cite{Horava:2005wm}. It would be interesting to study the transition between the $HO$ and $HE$ vacuum states from this perspective.

\section{Higher Dimensional String/M Theories}

Naturally we would like to generalize the approach of the previous sections to higher dimensional string theories. One approach discussed in section 2 is to use an asymmetric orbifold solution to heterotic string theory. This fixes the moduli to lie on a discrete lattice. One then has a large number of 4d vacuum with various cosmological constants to choose from \cite{Dienes:2006ut}. An approach to the cosmological seesaw would then involve taking two of these theories, one with $TeV^4$ scale vacuum energy and one with Planck scale  and allowing off diagonal transitions between these solutions.

Another approach is to generalize the above considerations of the 2d string cosmologies to higher dimensions. As the 2d string theories were $R \times S^1$ cosmologies we can consider string cosmologies of the form: $R\times S^2 \times S^1 \times K_6$, $R \times H^4 \times K_5$ etc where $K_n$ is an internal Kaluza-Klein space. One can also consider light cone M cosmology. We will consider each of these cases in turn.

\subsection{ $R \times \{T^2,S^2,H^2\} \times S^1 \times K_5$ Cosmology}

The first case we will consider is the space-time manifold $R \times \{T^2,S^2,H^2\} \times S^1 \times K_5$ which is a string generalization of the Kantowski-Sachs cosmology. The low energy effective action for this model is given by \cite{Cadoni:1994av}:
\[
S = \int {d^4 x\sqrt { - g} } e^{ - 2\phi } (R - \frac{{8 \alpha}}{{1 - \alpha}}(\nabla \phi )^2  - \frac{{3 + \alpha}}{{1 - \alpha}}F^2 )
\]
where the parameter $-1 \le \alpha  \le 1$ relates an internal  moduli field to the dilaton and $F$ is a $U(1)$ gauge field. We are mainly interested in the case $\alpha =-1$ which is closely related to the 2d string theories considered above.
 
Of particular interest is the space time $R\times S^2 \times S^1 \times K_6$ which is related to four dimensional black holes through a $r \leftrightarrow t$ double Wick rotation. For this case one can use the ansatz:
\[
\begin{array}{l}
 ds^2  =  - N^2 (t) + a^2 (t)d\Omega_{S^1} ^2  + b^2 (t)d\Omega _{S^2}^2  \\ 
 \phi  = \phi (t) \\ 
 F = Q_m \omega_{S^2}  \\ 
 \end{array}
\]
Here $d\Omega_{S^n}^2$ and $\omega_{S^n}$ are the line element and volume form on $S^n$ respectively. Substantial simplification occurs if the radius of the $S^2$ is fixed by the magnetic flux. We then have $b(t) = b_0  = Q_m \sqrt 2$ and the theory reduces to a 2d effective action given by \cite{Cadoni:1994av}:
\[
S = \int {d^2 x\sqrt { - g} } e^{ - 2\phi } (R + 4(\nabla \phi )^2  + \frac{1}{{b_0^2 }})
\]
The Friedmann equation is given by:
\[
 - 4\frac{{\dot a}}{a}\dot \phi  + 4\dot \phi ^2  + \frac{1}{{b_0^2 }} = 0
\]
and the Hamiltonian constraint by:
\[
 - \frac{{e^{2\phi } }}{4a}\pi _a \pi _\phi   - \frac{{e^{2\phi } }}{4}\pi _a^2  + e^{ - 2\phi } \frac{1}{{b_0^2 }} = 0
\]
The WDW equation is then:
\[
\left( {\frac{{e^{2\phi } }}{4a}\partial _a \partial _\phi   + \frac{{e^{2\phi } }}{4}\partial _a^2  + e^{ - 2\phi } \frac{1}{{b_0^2 }}} \right)\Psi (a,\phi ) = 0
\]
A large $N$ dual for this cosmology is not known but given the similarity with the 2d cases discussed in section 3 it should be closely related to the Matrix cosmologies studied by \cite{Karczmarek:2003pv}. When the $S^2$ radius is not fixed by the magnetic flux one has a more complicated String Kantowski-Sachs cosmology \cite{Cadoni:1999gh}\cite{Conradi:1996qu}\cite{Barrow:1996gx}\cite{Conradi:1994yy}\cite{Nambu:1987dh} where the wave function is of the form $\Psi(a,b,\phi)$. The large $N$ dual is expected to be more complicated as well. For a $R\times H^2 \times S^1 \times S^3 \times K_3$ cosmology one expects a large $N$ dual given by a two dimensional CFT.

\subsection{$R \times \{T^4, S^4, H^4\} \times K_5 $ Cosmology}

When $K_5 =S^5$ is a fixed internal space this has the interpretation of a 5d cosmology. In particular for $H^4$ spatial geometry this cosmology is constructed by double Wick rotation of the $AdS_5 \times S^5$ supersymmetric solution \cite{Bak:2006nh}. The time dependence is a way to parametrize a deformation of the solution. The ansatz for this spacetime is of the form:
\[
\begin{array}{l}
 ds^2  =  - N^2 (t)dt^2  + a^2 (t)d\Omega _{H^4 }^2  + b_0^2d\Omega _{S^5 }^2  \\ 
 \phi  = \phi (t) \\ 
 F_5  = 2a^4 (t)dt \wedge \omega_{H^4}  + 2b_0^5\omega_{S^5 }  \\ 
 \end{array}
\]
Here the five form flux $F_5$ goes through both the the $R\times H^4$ and the $S^5$ space. 

The analog of the Friedman equation for this spacetime is given by:
\[
4\left( {\frac{{\dot a}}{a}} \right)^2  =  - \frac{4}{{b_0^2 }} + \frac{4}{{a^2 }} + \frac{{\dot \phi ^2 }}{6}
\]
The dilaton can be expressed in terms of it's canonical momentum as:
\[
\pi _\phi   = a^4 \dot \phi  = c_0 
\]
And the Friedmann equation becomes:
\[
4\left( {\frac{{\dot a}}{a}} \right)^2  =  - \frac{4}{{b_0^2 }} + \frac{4}{{a^2 }} + \frac{{c_0^2 }}{{6a^8 }}
\]
The Hamiltonian constraint is then:
\[
\frac{1}{{24M_{Pl}^3 a^6 }}\pi _a^2  =  - \frac{{4M_{Pl}^3 }}{{b_0^2 }} + \frac{{4M_{Pl}^3 }}{{a^2 }} + \frac{{\pi _\phi ^2 M_{Pl}^3 }}{{6a^8 }}
\]
The WDW equation is:
\[
 - \frac{1}{{24M_{Pl}^3 a^6 }}\frac{{\partial ^2 }}{{\partial a^2 }}\Psi (a,\phi ) = ( - \frac{{4M_{Pl}^3 }}{{b_0^2 }} + \frac{{4M_{Pl}^3 }}{{a^2 }})\Psi (a,\phi ) - \frac{{M_{Pl}^3 }}{{6a^8 }}\frac{{\partial ^2 }}{{\partial \phi ^2 }}\Psi (a,\phi )
\]
Fourier transforming the wave function with respect to the dilaton canonical momentum we can write this as:
\[
 - \frac{1}{{24M_{Pl}^3 a^6 }}\frac{{\partial ^2 }}{{\partial a^2 }}\Psi _{\pi _\phi  } (a) = ( - \frac{{4M_{Pl}^3 }}{{b_0^2 }} + \frac{{4M_{Pl}^3 }}{{a^2 }})\Psi _{\pi _\phi  } (a) + \frac{{M_{Pl}^3 \pi _\phi ^2 }}{{6a^8 }}\Psi _{\pi _\phi  } (a)
\]
The large N dual of this model was proposed as \cite{Bak:2006nh}:
\[
S = \int {d^4 x\frac{{2}}{{g^2 (t)}}Tr\{ \frac{1}{4}} (F_{\mu \nu } )^2  + \frac{1}{2}(D_\mu  \phi _i )^2  - \frac{1}{4}[\phi _i ,\phi _j ][\phi _i ,\phi _j ] + \frac{1}{2}\bar \psi D\psi  - \frac{i}{2}\bar \psi \Gamma _i [\phi _i ,\psi ]\} 
\]
where $\phi_i$ are six bosonic fields and $\psi$ four Majorana fermions each in the adjoint of $SU(N)$. The novel feature is the time dependence of the coupling constant of the large N theory. Thus scale change in the dual large $N$ gauge theory corresponds to cosmological evolution in time.

This solution breaks supersymmetry so a nonzero cosmological constant is generated in the above theory. A supersymmetric version of the Janus solution also exists \cite{Clark:2005te}\cite{D'Hoker:2006uu}. In that case the cosmological constant would be zero. Mixing between these spacetimes would a natural setting to discuss the cosmological constant seesaw similar to the 2d case of section 3.

\subsection{Light Cone Cosmology}

In the case of Light Cone cosmology the spacetime has dependence on a light cone coordinate rather than time alone. This is particularly useful for M-theory cosmologies where the ansatz takes the form \cite{Craps:2005wd}:
\[
ds^2  =  - a^2 (U)ds_{10}^2  + b^2 (U)dY^2 
\]
From the string perspective the radius of the extra 11th dimension is described by a dilaton field that depends on the light cone coordinate $U$. In that case the ansatz becomes:
\[
\begin{array}{l}
 ds_{10}^2  =  - N^2 (U)dUdV + a^2 (U)\sum\limits_{i = 1}^8 {dX^{i2} }  \\ 
 \phi  = \phi (U) \\ 
 \end{array}
\]
This nice feature of the light cone cosmology is that it's holographic dual is much simpler than the $R \times H^4 \times S^5$ case. When the dilaton field is $\phi(U) = -QU$ it is given by the 2d Large $N$ gauge theory \cite{Craps:2005wd}:
\[
S = \frac{1}{{2\pi \ell _s^2 }}\int {d\tau d\sigma \sqrt { - g} tr} (\frac{1}{2}(DX^i )^2  +  g_s^2 \ell _s^4 \pi ^2 F^2  - \frac{1}{{4\pi ^2 g_s^2 \ell _s^4 }}[X^i ,X^j ]^2 
\]
\[ + \theta ^T D\theta  + \frac{1}{{2\pi g_s \ell _s^2 }}\theta ^T \gamma _i [X^i ,\theta ])
\]
with a 2d $R \times S^1$ spacetime:
\[
ds^2  = e^{2Q\tau } ( - d\tau ^2  + d\sigma ^2 ) =  - dt^2  + a^2 (t)d\sigma ^2 
\]
The important fact about these cosmologies from the point of view of the cosmological constant seesaw is that they break supersymmetry and thus generate a nonzero cosmological constant in the string/M-theory spacetime. 
This was very difficult to do in a Holographic version of the 10d 0A or 0B theories for example \cite{Banks:1999tr}. 

\section{Conclusion}

In this paper we have extended the discussion of the cosmological constant seesaw to string/M-theories. We have found that off diagonal contributions to the WDW equation (transitions between different $ \lambda$)  exist for 2d/3d versions of String/M-theory. In higher dimensions vacua with $TeV^4$ and Planckian vacuum energies exist and it is plausible that transitions between different lambda also exist. These are the prerequisites for the cosmological constant seesaw. As the cosmological constant seesaw relation $\lambda_- = \lambda_1^2/\lambda_2$ is well defined in any dimension we started in 2d/3d string M-theories. We studied the transition between $0A$ and $0B$ theories in 2d/3d String/M-Theory as an example. We also studied transitions from $HE$ to $HO$ theories in 2d using Wilson lines. All these can be considered $R \times S^1$ cosmologies. To generalize to higher dimensions we considered $R \times S^2 \times S^1 \times K_6$ cosmologies with magnetic charge which were similar to the 2d cases. We also studied $R\times H^4 \times S^5 $  cosmologies with nonzero five form charge and finally cosmologies depending on a light cone coordinate. The existence of dual theories for these cosmologies with spontaneously broken supersymmetry and nonzero vacuum energy may lead to transitions between theories with different vacuum energies in the higher dimensional case as well.

\section*{Acknowledgements}
I wish to thank C. Nappi for suggesting the papers of  M.~Cadoni and M.~Cavaglia.

\end{document}